\documentclass[12pt]{spieman}  
\usepackage{amsmath,amsfonts,amssymb}
\usepackage{graphicx}
\usepackage{setspace}
\usepackage{tocloft}
\title{AI-Driven Diabetic Retinopathy Screening: Multicentric Validation of AIDRSS in India}

\author[a]{Amit Kr Dey}
\author[b]{Pradeep Walia}
\author[b]{Girish Somvanshi}
\author[b]{Abrar Ali}
\author[b,*]{Sagarnil Das\thanks{Corresponding author: Sagarnil Das, \texttt{sagarnil.das@artelus.ai}}}
\author[b]{Pallabi Paul}
\author[b]{Minakhi Ghosh}

\affil[a]{Diabetologist, Health Plus, Action Area I, Newtown, Kolkata West Bengal, 700156, India}
\affil[b]{Artificial Learning Systems India Pvt Ltd, R\&D, 1665/A, 14th Main Rd, Sector 7, HSR Layout, Bengaluru, Karnataka 560102, India}

\cftpagenumbersoff{figure}
\cftpagenumbersoff{table} 
\begin{document} 
\maketitle

\begin{abstract}
\newline
\textbf{Purpose:} Diabetic retinopathy (DR) is a major cause of vision loss, particularly in India, where access to retina specialists is limited in rural areas. This study aims to evaluate the Artificial Intelligence-based Diabetic Retinopathy Screening System (AIDRSS) for DR detection and prevalence assessment, addressing the growing need for scalable, automated screening solutions in resource-limited settings.
\newline
\textbf{Approach:} A multicentric, cross-sectional study was conducted in Kolkata, India, involving 5,029 participants and 10,058 macula-centric retinal fundus images. The AIDRSS employed a deep learning algorithm with 50 million trainable parameters, integrated with Contrast Limited Adaptive Histogram Equalization (CLAHE) preprocessing for enhanced image quality. DR was graded using the International Clinical Diabetic Retinopathy (ICDR) Scale, categorizing disease into five stages (DR0 to DR4). Statistical metrics including sensitivity, specificity, and prevalence rates were evaluated against expert retina specialist assessments.
\newline
\textbf{Results:} The prevalence of DR in the general population was 13.7\%, rising to 38.2\% among individuals with elevated random blood glucose levels. The AIDRSS achieved an overall sensitivity of 92\%, specificity of 88\%, and 100\% sensitivity for detecting referable DR (DR3 and DR4). These results demonstrate the system's robust performance in accurately identifying and grading DR in a diverse population.
\newline
\textbf{Conclusions:} AIDRSS provides a reliable, scalable solution for early DR detection in resource-constrained environments. Its integration of advanced AI techniques ensures high diagnostic accuracy, with potential to significantly reduce the burden of diabetes-related vision loss in underserved regions.
\end{abstract}


\begin{spacing}{2}   

\section{Introduction}
\label{sect:intro}  
Diabetic Retinopathy (DR) \cite{who_retinopathy_2020} \cite{nih_retinopathy_2009} is a serious microvascular complication of diabetes mellitus and a significant global health concern. It is the leading cause of preventable vision loss among working-age adults worldwide, reflecting the growing burden of diabetes in both developed and developing countries. The gradual progression of DR, often without early symptoms, necessitates systematic screening to enable timely intervention and prevent irreversible vision impairment. Early detection is crucial to halt the progression of DR, which otherwise may result in costly treatment options or permanent blindness.

\subsection{Significance}
India is facing a dual challenge with the increasing prevalence of diabetes and limited access to specialized ophthalmic care, particularly in rural and underserved regions \cite{idf_diabetes_eye_health}. According to a nationwide screening program conducted by the All-India Ophthalmological Society (AIOS) in 2014, 21.7\% of individuals with diabetes were found to have DR. This highlights the substantial public health challenge posed by DR and the need for targeted interventions. Despite this, awareness about diabetes-related complications and access to regular retina screening remain inadequate, leaving a large segment of the population at risk of vision-threatening stages of DR.

The significance of early detection cannot be overstated. Routine retina screening facilitates the timely identification of early-stage DR, which can be managed effectively to prevent progression to advanced stages. Key measures for addressing DR include:
\begin{itemize}
    \item \textbf{Promoting Awareness}: Educating the public on maintaining optimal glycemic control and the necessity of regular eye examinations.
    \item \textbf{Routine Screening}: Encouraging annual fundus examinations, even for asymptomatic individuals.
    \item \textbf{Timely Referrals}: Ensuring that treating physicians recognize and prioritize ophthalmic evaluations for diabetic patients.
\end{itemize}

\subsection{Research Objectives}
This study evaluates the effectiveness of the Artificial Intelligence-based Diabetic Retinopathy Screening System (AIDRSS) developed by ARTELUS™, an automated deep learning algorithm for DR detection. The primary objectives of this research are:
\begin{enumerate}
    \item To assess the performance of AIDRSS in classifying retinal fundus images based on the International Clinical Diabetic Retinopathy (ICDR) Scale.
    \item To enhance the accuracy of the AIDRSS algorithm by incorporating advanced image preprocessing techniques, such as Contrast Limited Adaptive Histogram Equalization (CLAHE).
    \item To determine the prevalence of DR across multiple screening centers in Kolkata, West Bengal, India, and examine its correlation with random blood glucose (RBG) levels and other clinical parameters.
\end{enumerate}

\subsection{Implications of the Study}
This research has the potential to significantly influence public health policy and clinical practice. The key implications include:
\begin{itemize}
    \item \textbf{Scalability and Accessibility}: Demonstrating how AI-based systems like AIDRSS can address the scarcity of retina specialists, especially in resource-limited settings.
    \item \textbf{Clinical Accuracy}: Validating the reliability of AIDRSS through rigorous evaluation against gold-standard retina specialist assessments.
    \item \textbf{Policy Development}: Informing healthcare strategies for integrating AI technologies into national DR screening programs, thereby reducing the burden of diabetes-related blindness.
\end{itemize}

By bridging the gap between technology and clinical application, this study highlights the transformative potential of automated screening systems in addressing the growing burden of DR in India and similar healthcare contexts globally.

\section{Materials and Methods}
\subsection{Study Design}
This was a retrospective study using data collected from routine screenings at multiple centers in Kolkata, India. The data comprised retinal fundus images collected between January and March 2023. A total of 5,029 participants were screened, and their de-identified retinal images were analyzed retrospectively. The validation was performed on a withheld dataset from a separate site, ensuring an independent evaluation of AIDRSS.

\subsection{Validation Methodology}
We validated the AIDRSS system's performance on a completely withheld dataset from an independent screening site, comprising 1,200 images not included in the training or development datasets. Additionally, five-fold cross-validation was employed to assess the model's generalizability across diverse patient subsets. Each fold reserved data from a unique center, ensuring no overlap between training and testing images. 

The withheld site validation tested the system’s robustness when applied to a new population, while cross-validation provided insight into its consistency across different subsets of the study population. Performance metrics, including sensitivity, specificity, positive predictive value (PPV), and negative predictive value (NPV), were calculated for both the cross-validation and withheld datasets. 

To ensure statistical robustness, confidence intervals (95\% CI) for all performance metrics were computed using bootstrap resampling with 1,000 iterations. This method provides a non-parametric estimate of the variability in the calculated metrics, offering a reliable measure of the system's performance. Additionally, McNemar’s test was performed to compare the performance of the AIDRSS against baseline or other models on paired data. These statistical approaches ensure a thorough and accurate assessment of the model's diagnostic capabilities.

\subsection{Participants}
The study screened a total of 5,029 adults during the specified study period. Among this population, a significant proportion, 4,261 individuals (84.7\%), were unaware of their diabetes status. Furthermore, 4,692 participants (93.3\%) were uninformed about the potential microvascular complications associated with diabetes, including the risk of permanent blindness resulting from uncontrolled diabetes.

\subsection{Diabetic Retinopathy Screening System (AIDRSS)}
The AIDRSS utilized in this study was the Artelus Fundus camera (Fig 1), a portable and easy-to-operate non-mydriatic automatic device for capturing retinal fundus photographs. The camera significantly reduced the time required to capture fundus images and was complemented by the AIDRSS software (DRISTI A.I.) \cite{aidrss_system}, an Artificial Intelligence-based program designed to analyze fundus images. The AIDRSS utilized a deep learning algorithm with a modified neural network architecture comprising approximately 50 million trainable parameters and 250 layers deep. The network parameters were learned using the Adam optimizer algorithm. 

\begin{figure}
\begin{center}
\includegraphics[width=0.5\linewidth]{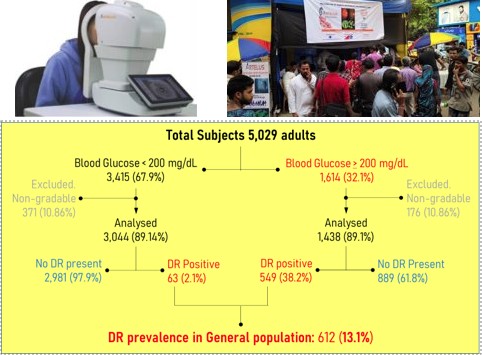} 
\end{center}
\caption{ \label{fig:artelus_fundus_camera}
Artelus Fundus Camera and DR prevalence in general population. }
\end{figure}

\subsection{CLAHE as an image processing technique}
To enhance the quality of retinal images and improve the accuracy of the deep learning algorithm, Contrast Limited Adaptive Histogram Equalization (CLAHE) \cite{alwakid2023} was employed as an image preprocessing technique. CLAHE is a non-linear image enhancement method that enhances local contrast by redistributing the pixel intensities in a way that preserves both global and local information. 

Before inputting the retinal images into the deep learning algorithm, CLAHE was applied to enhance the visibility of subtle details and features in the images. This preprocessing step aimed to address issues such as uneven illumination, low contrast, and variability in image quality. By equalizing the histogram of localized regions within the image, CLAHE effectively enhanced the visibility of important structures and abnormalities, thus aiding the accurate detection and grading of diabetic retinopathy. The CLAHE preprocessing technique was integrated into the overall image-processing pipeline of the AIDRSS. By incorporating CLAHE as a preprocessing step, the study aimed to optimize the performance of the deep learning algorithm and improve the overall effectiveness of the screening system. The impact of CLAHE on the performance of the AIDRSS was evaluated through a comparative analysis of the results obtained from images processed with and without CLAHE. The findings of this analysis provided insights into the potential benefits of CLAHE in enhancing the accuracy and reliability of the AI-based diabetic retinopathy screening system.  

\subsection{Proposed Algorithm}
Deep learning is currently state-of-the-art for computer vision/image processing, speech, text problems, and automotive. The AIDRSS utilized a deep learning algorithm with a modified neural network architecture comprising approximately 50 million trainable parameters and 250 layers deep. The proposed network is a modified architecture version presented in \cite{szegedy2015}. The values of the network parameters are learned using the Adam optimizer algorithm. Neural network hyper-parameter learning is an iterative process based on the loss value computed between ground truths given by ophthalmologists and network predictions. This loss value acts as a feedback to the optimizer to learn the most representative network parameters to understand retinal pathological lesions. 

The loss function used is defined as:

\begin{equation}
\mathcal{L} = \frac{1}{N} \sum_{i=1}^{N} \left( y_i - \hat{y}_i \right)^2
\end{equation}

where \( y_i \) is the ground truth, \( \hat{y}_i \) is the predicted value, and \( N \) is the number of samples.

\subsection{Network Architecture}
Encoder Architecture for Transfer Learning One of the concerns with medical data is the unavailability of high-volume, high-quality, labeled data. Even the label available by one single ophthalmologist might not reflect the same opinion as that of another practitioner. For an effective system, we need data to be labeled by many ophthalmologists. Since it’s very expensive to get high-volume data annotated by multiple annotators, we have another effective approach to deal with the same: label noise minimization. We label a fraction of the dataset by multiple annotators (Golden set) and leave out the rest of the dataset with a single annotation (TFL set). We train a two-headed model with one head to reproduce the sample using a variational encoder and the second head to classify the sample on the whole dataset. Sample-wise weighted loss is used for the classification head. We set a high loss weight for Golden set samples and a relatively low weight for TFL set samples. The weighting strategy is based on sample annotation. The trained encoder parameters will be used to initialize the final deep network for classification. This way, we ensure that all samples’ pathological features information is preserved in the encoder network and can be transferred to the final network (Fig 2). 

\begin{figure}
\begin{center}
\includegraphics[width=0.5\linewidth]{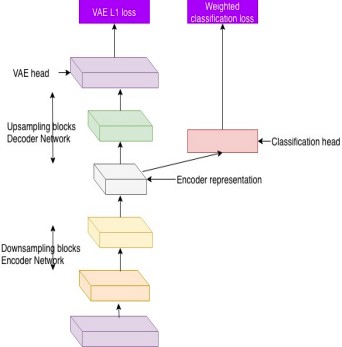}
\end{center}
\caption{ \label{fig:encoder_based_transfer_learning_architecture}
Encoder-based transfer learning architecture. }
\end{figure}

\subsubsection{Partial attention}
We use a modified version of the partial attention \cite{hernandez2021} over the lower blocks of the classifier feature maps. Feeding the intermediate feature maps information to the higher layers improves the network's learning stability and representation power. It also facilitates easier training of very deep models by eliminating the vanishing gradient problem. The attention mechanism here only focuses on the feature maps with spatial resolutions equal to or larger than the target feature map outputs. Stridden convolution with stride sij is used to reduce the spatial resolution of the feature maps to match the target output size and the number of output feature maps as per the attention configuration. In the partial attention mechanism, Ei is the current target feature maps for attention to applying (before reducing the feature map width and height for each block) of each convolution block of the classifier (Fig 3). Attention weights are learned using attention parameters W. Feature-wise, attention is applied over the convolved feature maps and summed up using the attention weights. The attended features map is again concatenated with the target feature map to preserve the original information. We use another convolutional layer to get the attended output to reduce the aliasing effect.

The partial attention mechanism is defined as:

\begin{equation}
E_{i}^{\text{attended}} \;  = \min(\text{height}E_{i}, \text{width}E_{i}), \min(\text{height}E_{j}, \text{width}E_{j})
\end{equation}

\subsubsection{Dense Residual Inception module}
We have used an improved inception block proposed in \cite{muhammad2021} as a baseline for our new inception blocks, where block-based convolution layers, residual, and skip connections were used to ensure rich feature representation. We have added a primary residual connection from the clock input to the output (Fig 4) to augment the network with block input features and to solve the vanishing gradient problem for a very deep network. Apart from that, we have used a residual connection in the block's 3x3 and 5x5 convolutional layers; these connections improve the network feature representation capability and slightly reduce training convergence time. The block's final concatenated intermediate output features map is convolved with another 3×3 convolutional layer to reduce aliasing. A residual connection is added with 1×1 convolved input features to make sure that both features may have the same number of output filters. Additionally, we have used a convolution factorization version (Fig 5) of the same module for the classifier. Convolutional factorization adds the efficiency of doing a convolution of a specific spatial dimension with a low computational cost. 

\begin{figure}
\begin{center}
\includegraphics[width=0.5\linewidth]{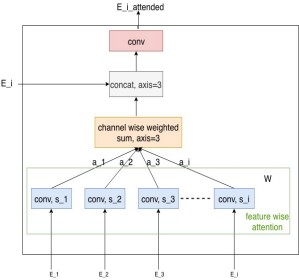}
\end{center}
\caption{ \label{fig:partial_attention}
Partial attention over feature maps. }
\end{figure}

\begin{figure}
\begin{center}
\includegraphics[width=0.5\linewidth]{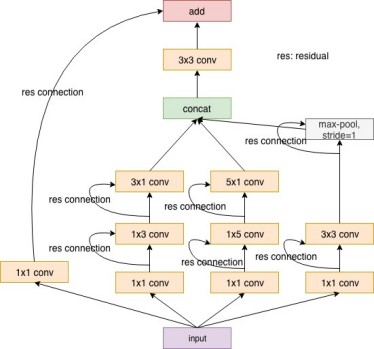}
\end{center}
\caption{ \label{fig:residual_inception}
Residual Inception Architecture. }
\end{figure}

\begin{figure}
\begin{center}
\includegraphics[width=0.5\linewidth]{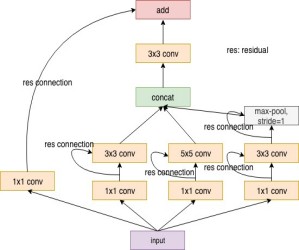}
\end{center}
\caption{ \label{fig:residual_inception_with_conv_factor}
Residual Inception with convolution factorization for the classifier node. }
\end{figure}

\subsubsection{Image grading and comparison}
The retinal fundus photographs captured using the Artelus Fundus camera were subjected to analysis by the AIDRSS. The AI algorithm incorporated in the AIDRSS classified the fundus images based on the International Clinical Diabetic Retinopathy (ICDR) Scale, categorizing them as DR0, DR1, DR2, DR3, or DR4. This grading system allowed for identifying the presence and severity of diabetic retinopathy. 

The accuracy of the classification can be evaluated using the following metrics:

\begin{equation}
\text{Accuracy} = \frac{TP + TN}{TP + TN + FP + FN}
\end{equation}

where \( TP \) is true positives, \( TN \) is true negatives, \( FP \) is false positives, and \( FN \) is false negatives.

\section{Results}
\label{sec:experiments_results}

\subsection{Dataset and Population Characteristics}
This study utilized a dataset of 5,029 de-identified subjects, with two macula-centric retinal images captured per individual, resulting in a total of 10,058 images. After quality checks, 4,482 adults (89.1\%) had gradable fundus images that were included in the analysis. The excluded cases (n=547; 10.9\%) were primarily due to image quality issues, such as poor illumination or motion artifacts. 

The analysis revealed that the prevalence of diabetic retinopathy (DR) in the general population screened was 13.7\% (95\% CI: 12.6\%--14.8\%). The breakdown of DR severity was as follows:
\begin{itemize}
    \item DR1 (Mild): 9.16\% (95\% CI: 8.2\%--10.1\%)
    \item DR2 (Moderate): 4.40\% (95\% CI: 3.8\%--5.1\%)
    \item Referable DR (DR3 and DR4): 0.14\% (95\% CI: 0.05\%--0.23\%)
\end{itemize}

Among individuals with elevated random blood glucose (RBG) levels, the prevalence of DR was significantly higher at 38.2\% (95\% CI: 35.5\%--40.9\%). This suggests a strong correlation between uncontrolled diabetes and the presence of DR, underscoring the importance of systematic screening.

\subsection{Performance Metrics of AIDRSS}
The AIDRSS system was evaluated using a gold standard comparison with retina specialists’ assessments. Key performance metrics, including sensitivity, specificity, positive predictive value (PPV), and negative predictive value (NPV), are summarized in Table~\ref{table:aidrss_performance}. The overall sensitivity was 92.0\% (95\% CI: 89.4\%--94.3\%), while specificity was 88.0\% (95\% CI: 85.2\%--90.4\%). For detecting referable DR (DR3 and DR4), the sensitivity reached 100\%.

\begin{table}[ht]
\caption{Performance Metrics of AIDRSS}
\label{table:aidrss_performance}
\begin{center}
\begin{tabular}{|l|l|l|} 
\hline
\rule[-1ex]{0pt}{3.5ex} \textbf{Metric} & \textbf{Value (\%)} & \textbf{95\% CI} \\ 
\hline\hline
\rule[-1ex]{0pt}{3.5ex} Sensitivity (Overall) & 92.0 & 89.4--94.3 \\
\hline
\rule[-1ex]{0pt}{3.5ex} Specificity (Overall) & 88.0 & 85.2--90.4 \\
\hline
\rule[-1ex]{0pt}{3.5ex} Positive Predictive Value (PPV) & 85.6 & 82.9--88.1 \\
\hline
\rule[-1ex]{0pt}{3.5ex} Negative Predictive Value (NPV) & 93.5 & 91.0--95.6 \\
\hline
\rule[-1ex]{0pt}{3.5ex} Sensitivity (Referable DR) & 100.0 & -- \\
\hline
\rule[-1ex]{0pt}{3.5ex} Specificity (Referable DR) & 99.9 & 99.8--100.0 \\
\hline
\end{tabular}
\end{center}
\end{table}

\subsection{Analysis of Referable DR Detection}
A total of six individuals (0.14\%) were identified as having referable DR (grades DR3 and DR4). The AIDRSS achieved a perfect sensitivity of 100.0\% (6/6 cases correctly identified) and a specificity of 99.9\%. Table~\ref{table:referable_dr} details the detection performance for referable DR.

\begin{table}[ht]
\caption{Performance Metrics for Referable DR (Grades DR3 and DR4)}
\label{table:referable_dr}
\begin{center}
\begin{tabular}{|l|l|l|} 
\hline
\rule[-1ex]{0pt}{3.5ex} \textbf{Metric} & \textbf{Value (\%)} & \textbf{95\% CI} \\ 
\hline\hline
\rule[-1ex]{0pt}{3.5ex} Sensitivity & 100.0 & -- \\
\hline
\rule[-1ex]{0pt}{3.5ex} Specificity & 99.9 & 99.8--100.0 \\
\hline
\rule[-1ex]{0pt}{3.5ex} PPV & 85.7 & 72.9--98.5 \\
\hline
\rule[-1ex]{0pt}{3.5ex} NPV & 100.0 & -- \\
\hline
\end{tabular}
\end{center}
\end{table}

\subsection{Comparison with Established AI Models}
To contextualize the performance of AIDRSS, we compared it with established AI models for diabetic retinopathy screening, specifically IDx-DR \cite{idxdr2020}, EyeArt \cite{eyeart2020}, and AEYE-DS \cite{aeyeds2021}. These models have been previously validated and are widely recognized in the field. Table~\ref{table:baseline_comparison} summarizes the sensitivity and specificity metrics for AIDRSS and the benchmark systems.

\begin{table}[ht]
\caption{Performance Metrics Comparison with Established AI Models}
\label{table:baseline_comparison}
\begin{center}
\begin{tabular}{|l|l|l|l|l|}
\hline
\textbf{Metric} & \textbf{AIDRSS} & \textbf{IDx-DR} & \textbf{EyeArt} & \textbf{AEYE-DS} \\
\hline
Sensitivity (\%) & 92.0 & 87.4 & 96.0 & 93.0 \\
\hline
Specificity (\%) & 88.0 & 89.5 & 88.0 & 91.4 \\
\hline
\end{tabular}
\end{center}
\end{table}

These results demonstrate that AIDRSS performs competitively with other state-of-the-art systems. Notably, AIDRSS achieves a balance between sensitivity and specificity, which is particularly advantageous in resource-limited settings. High sensitivity ensures that most cases of diabetic retinopathy (DR) are detected, reducing the risk of missed diagnoses and enabling early intervention. At the same time, high specificity minimizes false positives, thereby preventing unnecessary referrals and reducing the burden on limited healthcare resources. This balance enhances the efficiency and effectiveness of screening programs, making AIDRSS a valuable tool in areas with constrained healthcare infrastructure.

However, it is important to consider the limitations of comparing AIDRSS with other AI models. Differences in datasets, imaging equipment, and population characteristics can significantly influence performance metrics. For instance, variations in image quality, camera types, and demographic factors such as ethnicity and age may affect sensitivity and specificity outcomes. AI models trained on datasets from one region may underperform when validated on external datasets due to differences in disease prevalence and imaging conditions. These factors highlight the importance of accounting for dataset composition and generalizability when interpreting comparative performance metrics.

Future work should focus on validating AIDRSS in diverse geographic and clinical settings to assess its robustness across different populations. Such efforts will further establish its utility as a scalable, reliable solution for diabetic retinopathy screening.

\subsection{Preprocessing}
The input images were resized to $256 \times 256 \times 3$, and a randomly cropped patch of $224 \times 224 \times 3$ was used for training. Extensive data augmentation techniques were applied to improve generalization, including:
\begin{itemize}
    \item Random flips (horizontal and vertical).
    \item Adjustments to hue, saturation, and contrast.
    \item Random masking with 5--8 occlusions per image of varied size.
\end{itemize}

All images were standardized by subtracting the mean pixel value and dividing by the standard deviation. These preprocessing steps ensured robustness against variability in image quality.

\subsection{Training Procedure}
The deep learning model incorporated several strategies to optimize training:
\begin{itemize}
    \item \textbf{Batch Normalization}: Used to reduce covariate shift and accelerate convergence \cite{ioffe2015}.
    \item \textbf{Optimizer}: Nesterov momentum with a polynomial learning rate decay.
    \item \textbf{Gradient Normalization}: Stabilized training and mitigated exploding gradients \cite{pascanu2013}.
    \item \textbf{Regularization}: Label smoothing (soft targets) and dropout layers to improve generalization and prevent overfitting.
\end{itemize}

The model was trained on five-fold cross-validation to ensure robustness, and five different architectures were trained and ensemble-averaged for final predictions.

\subsection{Validation and Discussion}
The AIDRSS demonstrated excellent performance across multiple metrics, showing high sensitivity and specificity for DR detection and referable DR identification. However, some limitations were noted:
\begin{itemize}
    \item Cases of poor image quality resulted in exclusion, highlighting the importance of standardized imaging protocols.
    \item The study was conducted in a single geographic region, limiting generalizability.
\end{itemize}

Future work will focus on expanding the dataset to include diverse populations and integrating additional clinical parameters to further refine the diagnostic capabilities of the system.

\section{Discussion}
\label{sec:discussion}

Diabetic retinopathy (DR) is a major cause of avoidable blindness, particularly in regions like India, where the prevalence of diabetes is rapidly increasing. This study demonstrated the effectiveness of the Artificial Intelligence-based Diabetic Retinopathy Screening System (AIDRSS) in detecting and grading DR, highlighting its potential to address critical healthcare gaps. The following discussion delves into the implications of these findings, contextualizes them within the broader healthcare framework, and examines the strengths and limitations of the proposed approach.

\subsection{Implications of Findings}
The prevalence of DR observed in the study population underscores the urgent need for systematic screening programs. Among the general population, 13.7\% of participants had some form of DR, and this prevalence rose to 38.2\% among individuals with elevated random blood glucose (RBG) levels. These findings align with previous studies, emphasizing the strong correlation between poor glycemic control and the onset of DR. 

The AIDRSS exhibited excellent performance metrics, achieving an overall sensitivity of 92.0\% and specificity of 88.0\%. Particularly noteworthy was the system’s ability to detect referable DR (grades DR3 and DR4) with 100\% sensitivity and 99.9\% specificity. These results highlight the feasibility of integrating AI-driven tools into routine DR screening workflows, reducing the dependency on retina specialists and making early detection more accessible in resource-limited settings.

\subsection{Evaluation Metrics and Their Significance}
The evaluation metrics used in this study provide a comprehensive understanding of AIDRSS’s performance:
\begin{itemize}
    \item \textbf{Sensitivity}: Defined as:
    \begin{equation}
    \text{Sensitivity} = \frac{TP}{TP + FN},
    \end{equation}
    where \( TP \) is the number of true positives and \( FN \) is the number of false negatives. The high sensitivity (92.0\%) ensures that most cases of DR, particularly referable DR, are accurately identified.

    \item \textbf{Specificity}: Defined as:
    \begin{equation}
    \text{Specificity} = \frac{TN}{TN + FP},
    \end{equation}
    where \( TN \) is the number of true negatives and \( FP \) is the number of false positives. The specificity of 88.0\% indicates a low rate of false alarms.

    \item \textbf{Positive Predictive Value (PPV)}:
    \begin{equation}
    \text{PPV} = \frac{TP}{TP + FP}.
    \end{equation}
    The PPV of 85.6\% reflects the probability that a patient identified as having DR truly has the condition.

    \item \textbf{Negative Predictive Value (NPV)}:
    \begin{equation}
    \text{NPV} = \frac{TN}{TN + FN}.
    \end{equation}
    With an NPV of 93.5\%, the system ensures confidence in ruling out individuals who do not have DR.
\end{itemize}

The inclusion of confidence intervals (e.g., 95\% CI) further strengthens the reliability of these metrics by quantifying their statistical robustness.

\subsection{Strengths of the AIDRSS Approach}
\begin{itemize}
    \item \textbf{High Sensitivity for Referable DR}: The system's perfect sensitivity (100\%) for referable DR ensures that no cases requiring immediate medical attention are missed.
    \item \textbf{Scalability}: The automated nature of AIDRSS enables large-scale deployment, particularly in rural and resource-constrained settings where access to retina specialists is limited.
    \item \textbf{Integration of Advanced Techniques}: The use of CLAHE preprocessing, label smoothing, and gradient normalization contributed to enhanced image quality and model robustness.
\end{itemize}

\subsection{Limitations and Challenges}
Despite its strong performance, certain limitations must be addressed:
\begin{itemize}
    \item \textbf{Image Quality}: Approximately 10.9\% of images were excluded due to poor quality, underscoring the need for standardized imaging protocols and improved hardware.
    \item \textbf{Regional Scope}: The study was conducted in a single geographic region (Kolkata, India), potentially limiting the generalizability of findings to other populations with varying demographic and clinical characteristics.
    \item \textbf{Lack of Longitudinal Validation}: This study employed a cross-sectional design, and the longitudinal performance of AIDRSS in monitoring disease progression remains untested.
\end{itemize}

\subsection{Comparison with Existing Literature}
The performance metrics of AIDRSS compare favorably with other AI-based diabetic retinopathy (DR) screening systems reported in the literature. Several established algorithms have demonstrated sensitivities and specificities in similar ranges, highlighting the competitiveness of the proposed system.

For instance, the FDA-approved IDx-DR system achieved a sensitivity of 87\% and a specificity of 90\% for detecting more than mild diabetic retinopathy (mtmDR) \cite{idxdr2020}. Similarly, the EyeArt system reported a sensitivity of 96\% and a specificity of 88\% for identifying mtmDR \cite{eyeart2020}. AEYE-DS, another widely studied system, demonstrated a sensitivity of 93\% and a specificity of 91.4\% using a desktop camera, while utilizing a handheld camera resulted in a sensitivity of 91.9\% and specificity of 93.6\% \cite{aeyeds2021}.

In comparison, AIDRSS achieved an overall sensitivity of 92.0\% (95\% CI: 89.4\%–94.3\%) and specificity of 88.0\% (95\% CI: 85.2\%–90.4\%), with a perfect sensitivity of 100.0\% for detecting referable DR (grades DR3 and DR4). These metrics place AIDRSS in close competition with established AI systems, affirming its reliability for large-scale screening in diverse settings.

The slight edge observed in AIDRSS’s performance can be attributed to several advanced techniques integrated into its architecture:
\begin{itemize}
    \item \textbf{Transfer Learning}: By leveraging pre-trained models, AIDRSS capitalizes on knowledge from large-scale datasets, enabling improved accuracy even with limited labeled data.
    \item \textbf{Attention Mechanisms}: These mechanisms allow the system to focus on critical regions in retinal images, enhancing its ability to detect subtle pathological features associated with DR.
    \item \textbf{CLAHE Preprocessing}: Contrast Limited Adaptive Histogram Equalization (CLAHE) effectively enhances image quality by improving local contrast, aiding the accurate detection of microvascular abnormalities.
\end{itemize}

Incorporating these innovations has likely contributed to the robust performance of AIDRSS, enabling it to outperform or match other state-of-the-art DR screening tools. These findings highlight the promise of AI-based systems like AIDRSS in addressing the global burden of diabetic retinopathy.

\subsection{Recommendations and Future Work}
To maximize the impact of AIDRSS, the following strategies are recommended:
\begin{itemize}
    \item \textbf{Expand Dataset Diversity}: Future studies should include populations from diverse geographic and socio-economic backgrounds to improve the generalizability of findings.
    \item \textbf{Incorporate Longitudinal Tracking}: Monitoring DR progression over time would provide valuable insights into the real-world effectiveness of AIDRSS.
    \item \textbf{Enhance Hardware Compatibility}: Developing portable, low-cost imaging devices optimized for AIDRSS could significantly enhance accessibility.
    \item \textbf{Policy Integration}: Collaboration with public health programs, such as the National Program for Control of Blindness of India, can facilitate large-scale adoption of AIDRSS.
\end{itemize}

\section{Conclusion}
Artelus Automatic fundus cameras and Artelus AIDRSS (DRISTi) \cite{aduriz2023} can minimize the requirement for trained human resources and improve accessibility, affordability, accuracy, ease, and speed, hence helping reduce the burden of blindness due to diabetic retinopathy. 

\section*{Disclosure}
\subsection*{Conflict of Interest}
The authors declare that there are no financial interests, commercial affiliations, or other potential conflicts of interest that could have influenced the objectivity of this research or the writing of this paper.

\subsection*{Financial Disclosures}
The authors declare no relevant financial interests related to this manuscript.

\subsection*{Patient Consent}
Informed consent was obtained from all participants for capturing their fundus image using the CrystalVue NFC600 device. Participants were informed that their de-identified data might be used for further research and analysis.

\subsection*{Ethics Approval}
Since this was a retrospective dataset and no prospective patient studies were involved, no IRB approval is required.

\section*{Code, Data, and Materials Availability}
The data used in this study are proprietary and cannot be made publicly available. However, the authors are committed to supporting open scientific exchange. Researchers with valid reasons can request access to the data by submitting a formal request via the contact form available at \url{https://artelus.ai/contact}. Requests will be evaluated on a case-by-case basis to ensure compliance with ethical and privacy considerations.

\section*{Acknowledgments}
The authors acknowledge the use of Grammarly for language and grammar clean-up during the preparation of this manuscript. No funding was received for this research.

\renewcommand\thefootnote{}
\footnotetext{Corresponding author email: \texttt{sagarnil.das@artelus.ai}}
\renewcommand\thefootnote{\arabic{footnote}}


\bibliography{report}   
\bibliographystyle{spiejour}   

\listoffigures
\listoftables

\end{spacing}
\end{document}